\documentclass[aps,
reprint]{revtex4-2}

\usepackage{graphicx}
\usepackage{float}

\usepackage[utf8]{inputenc}
\usepackage{comment}
\usepackage{dcolumn}
\usepackage{bm}
\usepackage{caption}
\usepackage{subcaption}
\usepackage[boxed, vlined]{algorithm2e}

\usepackage{xcolor}
\usepackage{comment}

\usepackage{hyperref}
\usepackage{adjustbox}
\hypersetup{
    colorlinks=true,
    linkcolor=blue,
    filecolor=magenta,      
    urlcolor=cyan,
    pdftitle={Overleaf Example},
    pdfpagemode=FullScreen,
    }

\usepackage{natbib}
\usepackage{amsmath,amssymb}

\begin{document}

\title{Quantum circuits for simulating linear interferometers}
\author{Hudson Leone}
\affiliation{
     Centre for Quantum Software and Information -- University of Technology Sydney
     }
     \affiliation{Centre for quantum computation and Communication Technology (CQC2T)}
\author{Peter S.\ Turner}
 \email{peter.turner@mq.edu.au}
	\affiliation{School of Mathematical and Physical Sciences, Macquarie University, Sydney, New South Wales 2109, Australia}
    \affiliation{Sydney Quantum Academy, 1 Eddy Avenue, Sydney, New South Wales 2000, Australia}
\author{Simon Devitt}
 \email{Simon.Devitt@uts.edu.au}
     \affiliation{
     Centre for Quantum Software and Information -- University of Technology Sydney
     }
     \affiliation{InstituteQ, Aalto University, 02150 Espoo, Finland.}

\date{\today}

\begin{abstract}


Motivated by recent proposals for quantum proof of work protocols, we investigate approaches for simulating linear optical interferometers using digital quantum circuits.
We focus on a second quantisation approach, where the quantum computer's registers represent optical modes.
We can then use standard quantum optical techniques to decompose the unitary matrix describing an interferometer into an array of $2\times 2$ unitaries, which are subsequently synthesised into quantum circuits and stitched together to complete the circuit.
For an $m$ mode interferometer with $n$ identical photons, this method requires approximately $\mathcal{O}(m \log(n))$ qubits and a circuit depth of $\mathcal{O}(m n^4 \log_2(n) \: \textrm{polylog}(n^4 / \epsilon))$. 
We present a software package \texttt{Aquinas} (\textbf{a} \textbf{qu}antum \textbf{in}terferometer \textbf{as}sembler) that uses this approach to generate such quantum circuits.
For reference, an arbitrary five mode interferometer with two identical photons is compiled to a 10 qubit quantum circuit with a depth of 1972.

\end{abstract}

\maketitle

\section{Introduction}


Linear optics has been at the forefront of quantum information research for many years.
While interferometers natively execute certain types of high fidelity quantum operations, they cannot be practical universal quantum computers unless some nonlinearity is introduced \cite{Kok_07}.
Nevertheless there remains significant interest in linear quantum photonics, particularly on account of the seminal work of Aaronson and Arkhipov which demonstrated that the interferometry problem of \textit{boson sampling} is almost certainly
classically intractable \cite{Aaronson_11}.
This stems from the fact that even approximating the output occupation probabilities of a boson sampling experiment requires one to calculate the permanents of large matrices, which is a famously \#P-complete problem.

One implication of this is that linear optical interferometers, which are considerably easier to build than general purpose quantum computers, could in principle be used to demonstrate genuine (albeit impractical) quantum computational supremacy.
Significant work has therefore been done to engineer interferometers that are both large enough and reliable enough to demonstrate an incontestable quantum advantage (see Brod et.\ al.\ for a review of progress \cite{Brod_19}). 
Of note is a recent result from Madsen et.\ al.\ claiming supremacy using a variant of boson sampling that uses Gaussian states \cite{Madsen2022}.

Although linear interferometry lacks the computational power of a general purpose quantum computer, the proliferation of these experimental results has motivated research around finding practical applications for boson sampling.
Some computational examples include locating dense sub-graphs \cite{Arrazola_18} and the graph isomorphism problem \cite{Bradler_21}, though whether such algorithms can yield a tangible advantage in practice remains to be seen.
An arguably more promising application for boson sampling is in distributed consensus between quantum computers.
Recent work has indicated the suitability of coarse grained boson sampling as a proof of work algorithm, which may see use in future quantum infrastructure \cite{Singh_2024}.

Presently, to our knowledge the best reported algorithm for simulating linear interferometers is the classical method from Heurtel et.\ al.\ which, despite savings over conventional permanent based calculations, is still exponential in general (as expected) \cite{Heurtel2023}.
It is worth mentioning that a major practical issue with developing boson samplers is that of photon loss and distinguishability.
Though the hardness of boson sampling remains unchanged when a constant number of photons are lost \cite{Aaronson_16}, losing a fraction of the total number of photons can render the sampling problem classically tractable \cite{Oszmaniec2018, Moylett2019}.

One may then ask how to go about simulating boson sampling (and linear optics more generally) on a conventional quantum computer.
Of course any quantum computer that can emulate boson sampling will generally be more powerful than an interferometer.
Nevertheless, this could conceivably be useful for a quantum proof-of-work protocol, and simulating linear interferometers beyond what is experimentally feasible may be of fundamental interest.

In this paper we report a new software package called \texttt{Aquinas} (\textbf{A} \textbf{Qu}antum \textbf{In}terferometer \textbf{As}sembler) \cite{aquinas} which expresses a given linear interferometer as a quantum circuit that accurately simulates ideal linear interferometry for up to a pre-specified number of photons and modes. 
Broadly speaking, the algorithm works by first decomposing the given multimode interferometer into a network of interlaced two mode beamsplitters and phaseshifters.
Following this, the Hamiltonian of each two mode piece is expressed in terms of ladder operators over the respective modes.
By truncating the ladder operators at a fixed photon number (the total number of photons that enter the interferometer) and using some padding, we obtain a Hamiltonian that, when exponentiated, gives a unitary matrix that can be synthesised into a quantum sub-circuit.
These sub-circuits are then knitted together in the configuration of the original beamsplitters to form the interferometer circuit.

The main advantage of this `divide-and-conquer' approach is that, by bringing compilation down to the level of individual beamsplitters, the problem of synthesising an interferometer circuit becomes computationally tractable with respect to both $n$, the number of photons, and $m$, the number of modes.
As we will show, a clear drawback is that our circuits have an asymptotic depth that scales poorly with $n$.
Our technique is therefore best suited for sparse interferometers with many modes and a limited number of photons.
As a rule-of-thumb, a classically-hard instance of boson-sampling requires upwards of $n\geq30$ photons over $\mathcal{O}(n^2)$ modes \cite{Aaronson_11}. 
Our circuit depths become prohibitively large even for relatively easy experiments (as we will see in fig.\ \ref{fig:circuit_scaling}).
Nevertheless, we hope this work is a helpful first-step towards simulating quantum optics with digital quantum computers.


In the following section, we describe how a linear interferometer evolves the creation operators of a photon number state and explain how these states can be described in either first or second quantization. 
In section \ref{sec:simulating_second}, we limit our attention to the second quantization and detail our algorithm.
Finally, in section \ref{sec:results}, we present numerical analysis that indicates our software is working correctly while allowing us to approximate circuit depths for moderate instances of boson sampling.

\section{Linear optics and quantum computation background} \label{sec:background}

\begin{figure}[h]
    \centering
    \includegraphics[width=0.85\linewidth]{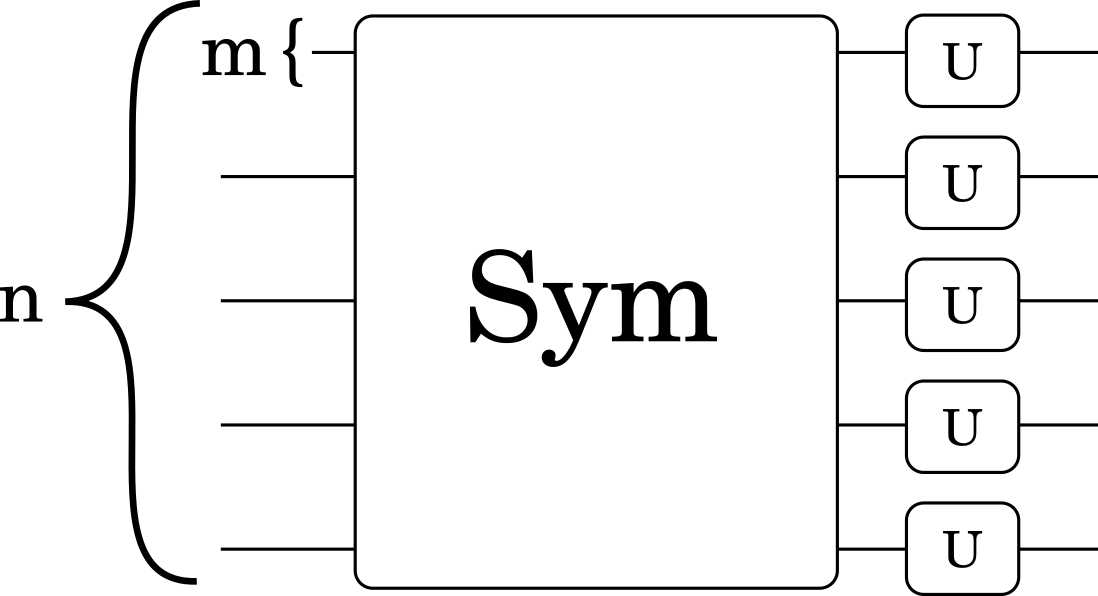}
    \caption{
    A generic quantum circuit for simulating a linear interferometer in first quantization.
    The $n$ registers are $m$-dimensional qudits that represent indistinguishable photons.
    Measurement in the computational basis indicates which of the $m$ possible modes that photon is found in.
    By the symmetrization postulate for bosons, reordering the $n$ registers can't change the input state; this is implemented by the highly entangling \textit{Sym} operation.
    Following this, the defining interferometer operation $U$ is applied to each photon locally.
    }
    \label{fig:first_quantized_circuit}
\end{figure}

Suppose we have a linear interferometer with $m$ input and $m$ output modes.
Such a device is usually characterised in second quantisation by an $m \times m$ unitary $U$ that describes how creation operators are transformed for each mode.
Specifically, if $\hat a^\dagger_i$ is the creation operator for the $i$th mode, we have
\begin{equation}
    \hat{a}^\dag_i \mapsto \sum_{i'=1}^m U_{ii'} \hat{a}^\dag_{i'} \label{eq:defU}
\end{equation}
where $U_{ii'}$ are the matrix elements of $U$.
For $n$ photons, the output state is described by a degree-$n$ polynomial in the creation operators.
The coefficients turn out to be permanents of $n\times n$ matrices $W$, whose matrix elements are taken from $U$ according to the input state and measurement outcome \cite{Scheel_08}





\begin{align}
    & \hat{a}^\dag_{i_1} \hat{a}^\dag_{i_2} \cdots \hat{a}^\dag_{i_n} \mapsto\nonumber\\ 
    &\left( \sum_{i'_1=1}^m U_{i_1 i'_1} \hat{a}^\dag_{i'_1} \right) \left( \sum_{i'_2=1}^m U_{i_2 i'_2} \hat{a}^\dag_{i'_2} \right) 
    \cdots \left( \sum_{i'_n=1}^m U_{i_n i'_n} \hat{a}^\dag_{i'_n} \right) \\
    &= \sum_{i_1' \leq i_2' \leq \cdots \leq i_n'} \mathrm{per} W^{i_1 i_2 \cdots i_n}_{i_1' i_2' \cdots i_n'}(U) \, \hat{a}^\dag_{i_1'} \hat{a}^\dag_{i_2'} \cdots \hat{a}^\dag_{i_n'}.
\end{align}

Quantum computation lends itself naturally to a first quantisation picture, where a simulation circuit for such a linear interferometer consists of $n$ quantum registers of dimension $m$.
Thus each register represents a photon, and measuring a register is analogous to learning the mode of that particular photon (see fig.\ \ref{fig:first_quantized_circuit})
\cite{Moylett_18}.
Because these particles are indistinguishable, we require by the symmetrization postulate that our input state is invariant under photon permutation.
Therefore the input state of our circuit must be invariant under permutation of the registers.
This means that the initial state must first be symmetrized, which is in general a non-trivial subroutine.
One could implement this symmetrization on a quantum computer with, for example, the Schur transform, which is known to be efficiently implementable \cite{Bacon_06}.

After symmetrization however, simulating the interferometer is straightforward because photon interactions are extremely weak and can be safely ignored.
Consequently, we can treat each photon of our symmetric state as if it were the only one passing through the interferometer.
Since $U$ can be interpreted as the transformation of a single photon,
the final step for simulating in first quantization is to implement $U$ on each of the $n$ registers.
A basic schematic of this protocol is presented in Fig.\ \ref{fig:first_quantized_circuit}.

\begin{figure}[h]
    \centering
    \includegraphics[width=0.75\linewidth]{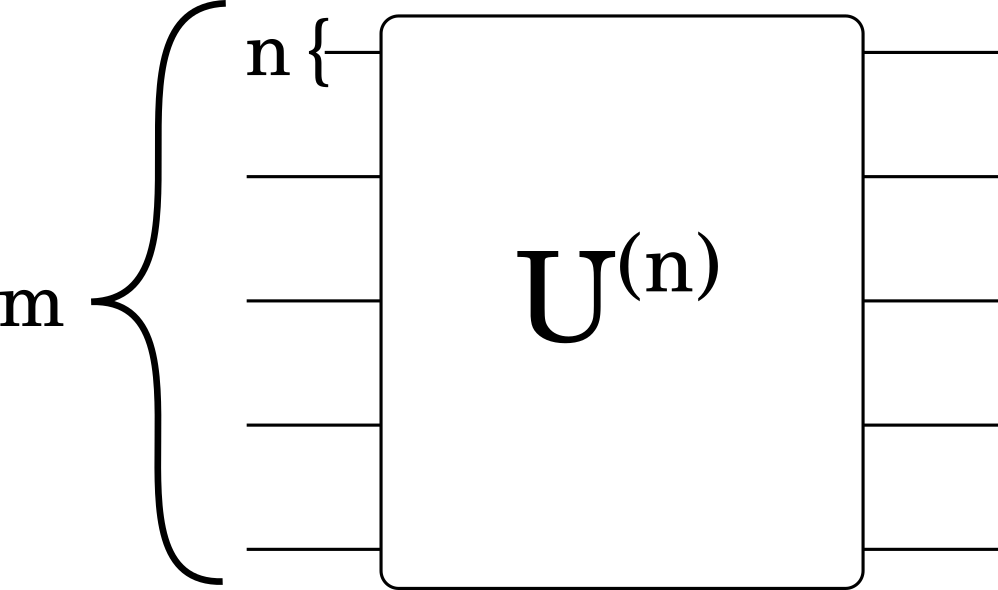}
    \caption{
    A generic quantum circuit for simulating a linear interferometer in the second quantization picture.
    The $m$ registers of the circuit are $n+1$ dimensional qudits that represent the $m$ modes of an interferometer.
    Measuring in the computational basis reveals how many photons are present in that mode.
    The complexity of the second quantization approach is in constructing a circuit to implement $U^{(n)}$, which is the unitary matrix corresponding to the interferometer $U$ acting in Fock space.
    }
    \label{fig:second_quantization_circuit}
\end{figure}

In second quantization, the roles of particles and modes are reversed (see fig.\ \ref{fig:second_quantization_circuit}).
Now we have $m$ registers of dimension $n+1$ that represent each of the possible modes.
Each mode can be measured to detect up to $n$ photons with an additional promise that there are $n$ photons in total.
Unlike the first quantization, there is no symmetrization step required in this picture since the modes only need to encode the number of particles present (the occupation).
The difficulty of simulating in second quantization is constructing a circuit to implement $U^{(n)}$, 
which is the unitary describing how photon number states with up to $n$ photons are transformed by the interferometer.
This is the problem we consider in this paper.

\section{Simulating in the second quantization} \label{sec:simulating_second}


We recall from fig.\ \ref{fig:second_quantization_circuit} that an optical simulation in the second quantization picture is realised with a quantum circuit that implements $U^{(n)}$; this unitary describes how photon number states containing up to $n$ photons evolve through the interferometer.
The primary difficulty of working in this quantization is finding a suitable way to construct this circuit.
To make headway into this problem, let us first consider how $U^{(n)}$ is related to $U$.
%
%
%
We begin by finding the Hamiltonian operator that generates $U$ under exponentiation
\begin{equation} \label{eq:hamiltonian}
    H = -i \; \textrm{log} \; U .
\end{equation}
Here, the above expression refers to the matrix logarithm.
It is well known that such a Hamiltonian is a linear combination of number preserving terms quadratic in the ladder operators (i.e. these terms generate the Lie algebra for the group of interferometer unitaries).

For a fixed photon number $n$, each mode need only support the $n+1$ occupations $0, 1, 2, \cdots, n$.
We therefore truncate the ladder operators, taking them from infinite dimensional to $n+1$ dimensional matrices 
\begin{equation}
    H \xrightarrow{\hat a_i \rightarrow \; \hat a_i^{(n)}} H^{(n)} .
\end{equation}
Finally, we exponentiate to obtain the truncated unitary
\begin{equation} \label{eq:U_truncd}
    U^{(n)} = e^{i H^{(n)}} .
\end{equation}


There are at least two approaches for efficiently constructing a circuit for $U^{(n)}$.
The first is Hamiltonian simulation, where the truncated Hamiltonian $H^{(n)}$ is decomposed into a polynomial-length sum of Pauli strings that are each `circuitized' and repeated to approximate the target unitary $U^{(n)}$.
Sawaya et.\ al.\ detail this approach in the context of quantum optics \cite{Sawaya2020}.
An alternative method, which is the one we present in this paper, is a `divide-and-conquer' approach wherein a standard decomposition of the interferometer \cite{Clements_16} is used to break $U$ into a grid of $2 \times 2$ unitaries (i.e. beamsplitters, or two mode interferometers) which have the form:
\begin{equation} \label{eq:clements_unitary}
U_{BS} = 
    \begin{pmatrix}
        e^{\phi} \cos \theta & -\sin \theta \\
        e^{i \phi} \sin \theta & \cos \theta
    \end{pmatrix}.
\end{equation}
Here, $\theta \in [0, \pi/2]$ is a parameter related to the \textit{reflectivity} of the beamsplitter; The device is transparent when $\theta=0$ and is perfectly reflective when $\theta=\pi/2$.
The $\phi \in [0, 2\pi]$ parameter is the relative phase shift introduced between the output modes.

\begin{figure*}
    \centering
    \includegraphics[width=0.8\linewidth]{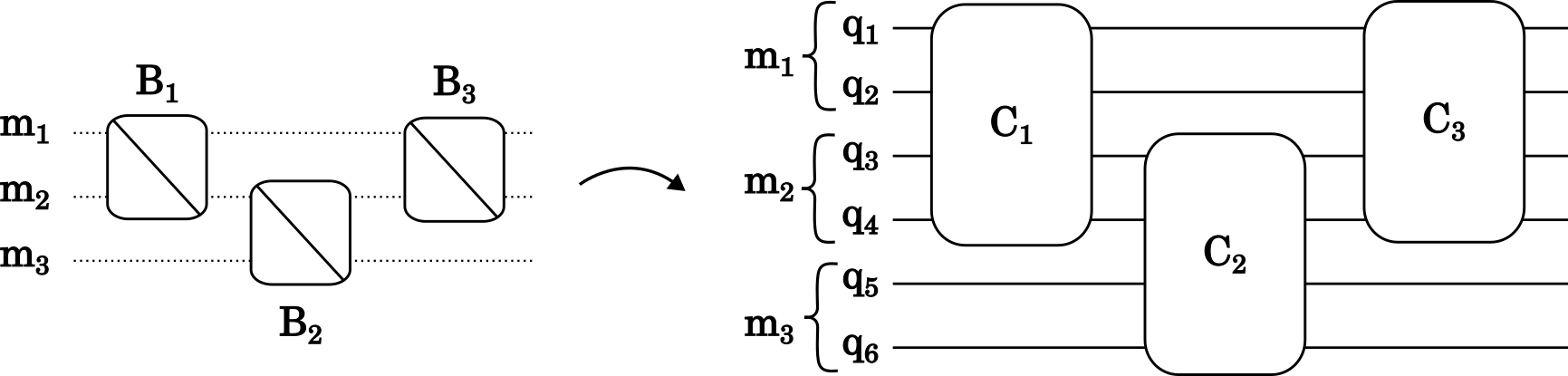}
    \caption{An illustration of how the `divide-and-conquer' method translates a 3 mode interferometer into a quantum circuit. (Left): According to the Clements decomposition, an arbitrary linear interferometer with 3 spatial modes ($m_1$, $m_2$, $m_3$) contains three beamsplitter elements ($B_1, B_2, B_3$). (Right): Assuming a photon number $n<4$, two qubits can be used to encode the photon number at each of the three modes. Qubits ($q_1, q_2$) for example encode the photon number at mode $m_1$. Individual beamsplitters are translated into quantum circuits ($C_1, C_2, C_3$) that act between the respective modes. Here, the $B_i$ beamsplitter corresponds to the $C_i$ circuit. The average circuit depth and CNOT count for each $C_i$ in this example is around 189 and 97 respectively (these values were calculated by first sampling 100 random beamsplitters and building the corresponding circuits given $n=3$).
    }
    \label{fig:circuit_illustration}
\end{figure*}

To calculate the truncated counterparts of these unitaries, we follow the steps laid out in equations \ref{eq:hamiltonian} through \ref{eq:U_truncd}.
After synthesising circuits for each of these simpler unitaries, we can take matrix products according to their original configuration to obtain a circuit for $U^{(n)}$ (see fig.\ \ref{fig:beamsplitter_decomp} for a schematic of the decomposition and fig.\ \ref{fig:circuit_illustration} for an illustrative example of the basic circuit structure).


\subsubsection{Two mode interferometer}
\label{sec:worked_example}

Before we analyze this approach numerically, let us illustrate with an example of how a truncated beamsplitter unitary is calculated.
Although \texttt{Aquinas} uses the Clements decomposition to obtain a grid of $2 \times 2$ unitaries in the form of eq.\ \ref{eq:clements_unitary}, we simplify our example by considering a different (but physically equivalent) beamsplitter unitary \footnote{The reason this unitary was chosen is because it can be expressed as a time-independent Hamiltonian. It is worth noting though that the Clements unitary given in eq.\ \ref{eq:clements_unitary} can be written as a \textit{product} of time-independent Hamiltonians (See eq.\ 4.13 in ref \cite{Arrazola2020}). This is needlessly complex for the sake of our example, which is why we use the eq.\ \ref{eq:simple_unitary} unitary instead.
}:

\begin{equation} \label{eq:simple_unitary}
    U_{BS} = 
    \begin{pmatrix}
        \cos \theta & e^{i \phi} \sin \theta \\
        -e^{-i \phi} \sin \theta & \cos \theta
    \end{pmatrix}. 
\end{equation}

The corresponding Hamiltonian is then:

\begin{equation}
    H_{BS} = -i \log U_{BS} = 
    \begin{pmatrix}
        0 & -i e^{i \phi} \theta \\
        i e^{-i \phi} \theta & 0
    \end{pmatrix}
\end{equation}
Which, written in terms of two-mode ladder operators, has the form:

\begin{equation} \label{eq:two_mode_ham}
    H_{BS} = 
    -i e^{i \phi} \theta \: \hat a_1^\dagger \hat a_2 + i e^{-i \phi} \theta \: \hat a_2^\dagger \hat a_1
\end{equation}

Say for example we want to find the corresponding unitary with with photon number $n = 2$.
In this case we would perform the following substitutions in equation \ref{eq:two_mode_ham}

\begin{equation} \label{eq:sub1}
    \hat a_1^\dagger \hat a_2 \rightarrow
    \begin{pmatrix}
        0 & 0 & 0 \\
        1 & 0 & 0 \\
        0 & \sqrt{2} & 0
    \end{pmatrix}
    \otimes
    \begin{pmatrix}
        0 & 1 & 0 \\
        0 & 0 & \sqrt{2} \\
        0 & 0 & 0 
    \end{pmatrix}
\end{equation}

\begin{equation} \label{eq:sub2}
    \hat a_2^\dagger \hat a_1 \rightarrow
    \begin{pmatrix}
        0 & 1 & 0 \\
        0 & 0 & \sqrt{2} \\
        0 & 0 & 0 
    \end{pmatrix}
    \otimes
    \begin{pmatrix}
        0 & 0 & 0 \\
        1 & 0 & 0 \\
        0 & \sqrt{2} & 0
    \end{pmatrix}
\end{equation}
and then take the matrix exponential to obtain $U_{BS}^{(2)}$:


\begin{widetext}
\begin{equation}
\resizebox{1.\textwidth}{!}{$
\begin{pmatrix}
 1 & 0 & 0 & 0 & 0 & 0 & 0 & 0 & 0 \\
 0 & \text{Cos}[\theta ] & 0 & -e^{-i \phi } \text{Sin}[\theta ] & 0 & 0 & 0 & 0 & 0 \\
 0 & 0 & \text{Cos}[\theta ]^2 & 0 & -\sqrt{2} e^{-i \phi } \text{Cos}[\theta ] \text{Sin}[\theta ] & 0 & e^{-2 i \phi } \text{Sin}[\theta ]^2 &
0 & 0 \\
 0 & e^{i \phi } \text{Sin}[\theta ] & 0 & \text{Cos}[\theta ] & 0 & 0 & 0 & 0 & 0 \\
 0 & 0 & \sqrt{2} e^{i \phi } \text{Cos}[\theta ] \text{Sin}[\theta ] & 0 & \text{Cos}[2 \theta ] & 0 & -\sqrt{2} e^{-i \phi } \text{Cos}[\theta
] \text{Sin}[\theta ] & 0 & 0 \\
 0 & 0 & 0 & 0 & 0 & \text{Cos}[2 \theta ] & 0 & -2 e^{-i \phi } \text{Cos}[\theta ] \text{Sin}[\theta ] & 0 \\
 0 & 0 & e^{2 i \phi } \text{Sin}[\theta ]^2 & 0 & \sqrt{2} e^{i \phi } \text{Cos}[\theta ] \text{Sin}[\theta ] & 0 & \text{Cos}[\theta ]^2 & 0 &
0 \\
 0 & 0 & 0 & 0 & 0 & e^{i \phi } \text{Sin}[2 \theta ] & 0 & \text{Cos}[2 \theta ] & 0 \\
 0 & 0 & 0 & 0 & 0 & 0 & 0 & 0 & 1 \\
\end{pmatrix}
$}
\end{equation}
\end{widetext}
Notice however that the size of this matrix is not equal to a power of two since the creation and annihilation operators we substituted in equations \ref{eq:sub1} and \ref{eq:sub2} were not themselves powers of two.
We have two options for expanding into a qubit-friendly Hilbert space.
The first and most obvious choice is to pick the next highest truncation that has a power-two dimension.
In this example, that would be $n=3$

\begin{equation}
    \hat a^{\dagger (3)} = 
    \begin{pmatrix}
        0 & 0 & 0 & 0 \\
        1 & 0 & 0 & 0 \\
        0 & \sqrt{2} & 0 & 0 \\
        0 & 0 & \sqrt{3} & 0
    \end{pmatrix}
\end{equation}

Alternatively, if one favours a slightly more sparse matrix, one can simply \textit{pad} the creation, annihilation, and identity operators with zeros until the next power of two is reached.
For example,

\begin{equation} \label{eq:padding}
    \hat a^{\dagger (2)}_{\textrm{padded}} = 
    \begin{pmatrix}
        0 & 0 & 0 & 0 \\
        1 & 0 & 0 & 0 \\
        0 & \sqrt{2} & 0 & 0 \\
        0 & 0 & 0 & 0
    \end{pmatrix}
\end{equation}

To see the effect of this padding, consider a padded operator of the form $\hat a^\dagger_i \hat a_j$. 
When exponentiated, the action of the resulting unitary is seen with some effort to be idempotent on states with a photon number greater than the truncation depth.





\subsection{Analysis of `divide and conquer'} \label{sec:divide_and_conquer}




At the start of section \ref{sec:simulating_second}, we  illustrated how the Clements decomposition can be used to decompose any interferometer into a grid of two-mode interferometers.
Synthesising circuits for each of these elements is significantly easier than finding a circuit for $U^{(n)}$ since fewer `active' qubits are required.
To see how many exactly, first note that the minimum number of qubits needed to encode up to $n$ photons in a single mode is $\lceil \log_2(n+1) \rceil$.
A two mode unitary truncated at a photon number of $n$ therefore acts on $2 \lceil \log_2(n+1) \rceil$ qubits.
Recall that by the Solovay-Kitaev theorem, the number of gates required for a generic $k$ qubit operator scales as \cite[Chapter 4, Section 5.4]{Nielsen2012}
\begin{equation} \label{eq:solovay}
    \mathcal{O}\bigg(4^k k^2 \log ^c(4^k k^2 / \epsilon)\bigg)
\end{equation}
Where $\epsilon$ is the desired accuracy and $c$ is a constant.

To find the asymptotic depth of a beamsplitter circuit with respect to the photon number $n$, we make the approximation $k \approx 2 \log_2(n)$ and assume that the gates are uniformly distributed across the wires. The asymptotic circuit depth is therefore equal to the number of gates divided by the number of qubits.

\begin{equation}
    \mathcal{O}\Big(n^4 \log_2(n) \: \log^c(2 n^4 \log_2^2(n)/\epsilon)\Big)
\end{equation}

which, to be succinct, we write as
\begin{equation}
    \mathcal{O}\Big(n^4 \log_2(n) \: \textrm{polylog}(n^4 / \epsilon) \Big).
\end{equation}


Although the depth of a beamsplitter circuit is evidently polynomial with respect to $n$, it remains to be seen whether the \textit{compilation complexity} is polynomial (in other words, can these circuits be efficiently constructed with respect to $n$?).
To answer this question, it is sufficient to consider how the size of the Hilbert space grows with $n$.  Because $2\lceil \log_2(n+1) \rceil$ qubits are required to encode up to $n$ photons per mode over two modes, the dimension of the 'active' beamsplitter Hilbert space is
\begin{equation}
    \textrm{dim}(\mathcal{H}) = 2^{2\lceil \log_2(n+1) \rceil}
\end{equation}
which is asymptotically equivalent to $\mathcal{O}(n^2)$. Since the size of the Hilbert space grows quadratically with respect to $n$, we conclude that circuit synthesis is tractable for each beamsplitter element. 


Having established the width and depth of a \textit{single} beamsplitter circuit we now consider the number of beamsplitters that are required to decompose the interferometer. 
Well established results in mathematics \cite{Hurwitz1897} and more recently in the context of quantum optics \cite{Reck_94, Clements_16} show that $m (m-1) / 2$ two-mode interferometers are required to reconstruct an arbitrary $m$-mode interferometer, which immediately implies a depth of $\mathcal{O}(m)$ beamsplitter elements and therefore a total circuit depth of $\mathcal{O}(m n^4 \log_2(n) \: \textrm{polylog}(n^4 / \epsilon) )$.
In summary, the required number of qubits and the circuit depth of this second quantization approach are:

\renewcommand{\arraystretch}{1.8}
\begin{center}
\begin{tabular}{|c|c|}
  \hline
  Size & $\mathcal{O}(m \log(n))$ \\ \hline
  Depth & $\mathcal{O}\Big(m n^4 \log_2(n) \: \textrm{polylog}(n^4 / \epsilon)\Big)$ \\ \hline
\end{tabular}
\end{center}
\renewcommand{\arraystretch}{1.0}

Moreover, since the number of beamsplitters in a given interferometer decomposition is $\mathcal{O}(m^2)$, it follows that our divide-and-conquer approach also has a polynomial compilation complexity with respect to $m$.

\begin{figure}
    \centering
    \includegraphics[width=1.0\linewidth]{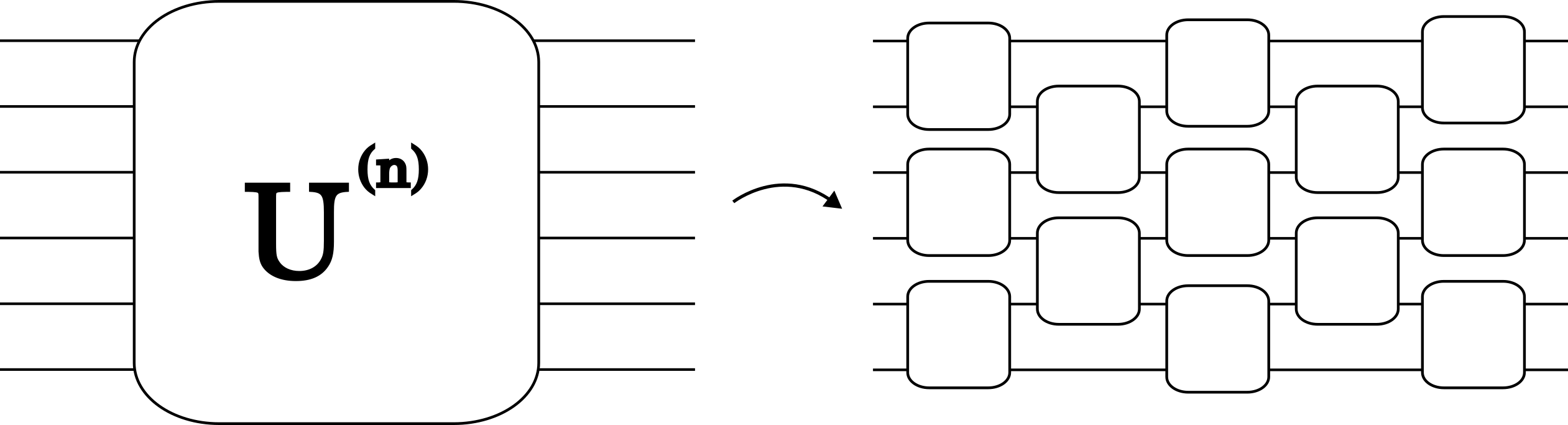}
    \caption{Given a linear interferometer $U$, it is not computationally tractable to generate a circuit for the truncated operator $U^{(n)}$.
    In the `divide-and-conquer' approach, we use the decomposition of \cite{Clements_16} to break $U$ into a grid of two-by-two unitaries that are significantly smaller than the original interferometer matrix.
    As such, this method is best suited for simulating sparse linear interferometery where $n \ll m$, as in boson sampling.
    }
    \label{fig:beamsplitter_decomp}
\end{figure}

\section{Results} \label{sec:results}

The \texttt{Aquinas} software package \cite{aquinas} takes an $m \times m$ unitary matrix representing an $m$ mode linear interferometer and uses the method described in section \ref{sec:divide_and_conquer} to synthesise a quantum circuit that simulates the interferometer for any input state of up to $n$ photons. 
To compensate for the fact that the dimensions of creation and annihilation operators truncated at photon number $n$ generally aren't a power of two, we use the operator padding technique described in the worked example of section \ref{sec:simulating_second} to ensure that $U^{(n)}$ can be implemented as a quantum circuit.
(See algorithm \ref{alg:aquinas} for pseudo-code)

Our method is perhaps best described by way of examples -- we present two numerically simulated interferometer experiments.
In the first of these, we randomly selected a three mode interferometer $U_{\textrm{rand}} \in \textrm{SU}(3)$ and began with three indistinguishable photons in the arbitrarily chosen configuration $(2, 1, 0)$ (which is to say there are two photons in the first mode and one in the second mode).
Using \texttt{Aquinas}, we then constructed a quantum circuit that implements the transformation $U^{(3)}_{\textrm{rand}}$ using 6 qubits with a depth of 567.
We simulated this circuit noiselessly using the \texttt{Qiskit} \cite{qiskit2024} software package to obtain 10,000 measurement samples which were then converted into estimates for the probabilities of observing each of the 10 possible output configurations.
In order to validate these probabilities, we used the permanent method to the calculate the expectation values for each outcome analytically.

\begin{figure*}[tb]
    \centering
    \begin{subfigure}{0.60\textwidth}
         \centering
         \includegraphics[width=1.0\linewidth]{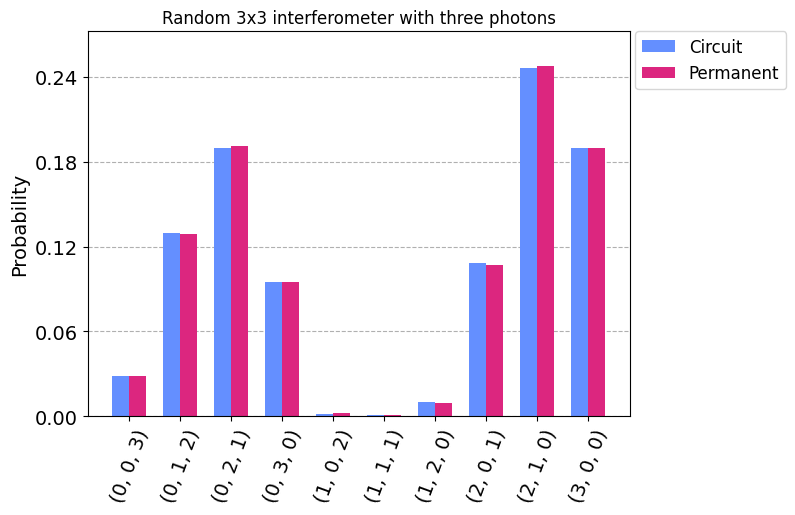}
         \label{fig:1a}
         \caption{A randomly selected $3 \times 3$ interferometer with up to three photons
         is simulated as a 6 qubit, depth 567 circuit.
         The initial photon configuration (2,1,0) is scrambled into one of 10 possible output distributions.
         }
    \end{subfigure}
    \begin{subfigure}{0.60\textwidth}
        \includegraphics[width=1.0\linewidth]{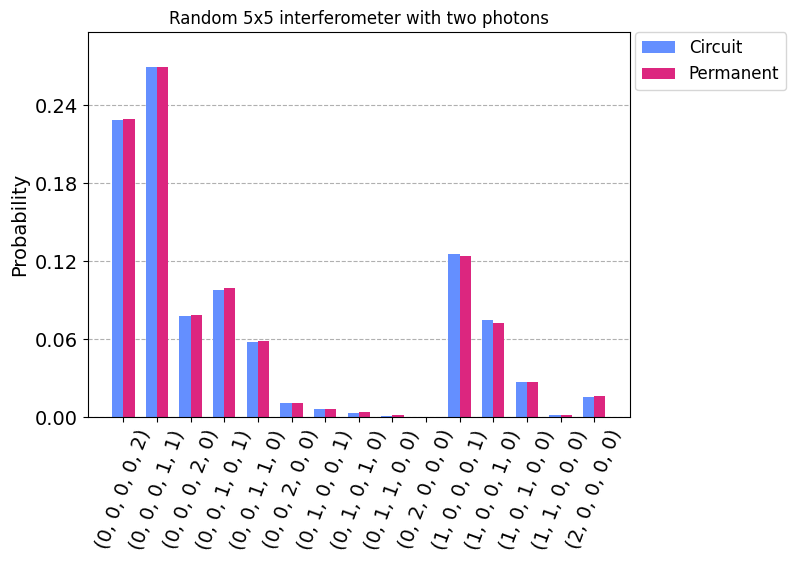}
        \caption{A randomly selected $5 \times 5$ interferometer with up to two photons is simulated as a 10 qubit, depth 1972 circuit.
        The initial photon configuration (2,0,0,0,0) is scrambled into 14 possible output distributions.
        }
        \label{fig:1b}
    \end{subfigure}
    \caption{
    Two numerical experiments in which the output statistics of interferometer circuits are compared with genuine expectation values (calculated via the matrix permanent method).
    The circuit statistics in blue were compiled by running each circuit 10,000 times.
    }
    \label{fig:random_3x3}
\end{figure*}

We compare the estimated probabilities with the exact expectation values for this experiment using the histogram in the upper part of figure \ref{fig:random_3x3}.
From inspection, it appears that our estimates converge on the genuine expectation values with minor variations that may possibly be attributed to sampling uncertainty.
To verify that this convergence was genuine, we began by using \texttt{qiskit}'s built-in \texttt{Operator} class to transform the circuit into a $2^6 \times 2^6$ matrix.
After this, we calculated the expectation values of a random assortment of output probabilities by `looking up' the appropriate matrix element:

\begin{equation}
    p_{\textrm{out}} = \Big|\langle \psi_\textrm{out}| U_{\textrm{rand}}^{(3)} |\psi_\textrm{out} \rangle\Big|^2
\end{equation}
With this, we were able to quickly verify that the expectation values of the circuit were accurate to an error of around $\epsilon \approx 3 \times 10^{-15}$.

The second experiment was virtually identical to the first, with the only difference being that we instead simulated a five mode interferometer with two photons in the starting configuration $(2,0,0,0,0)$.
In this case, the resulting 10 qubit circuit has a depth of 1972.
The estimated probabilities of the 15 possible output distributions from $10,000$ circuit samples are plotted in the bottom part of figure \ref{fig:random_3x3} together with the analytical expectation values.
As before, we see convincing evidence of genuine convergence that we confirmed by manually checking the expectation values of the operator corresponding to the circuit.
As before, we compared the expectation values of this circuit to the analytical values and found a somewhat higher error of around $\epsilon \approx 6 \times 10^{-15}$.

\begin{figure*}
    \centering
    \includegraphics[width=0.50\linewidth]{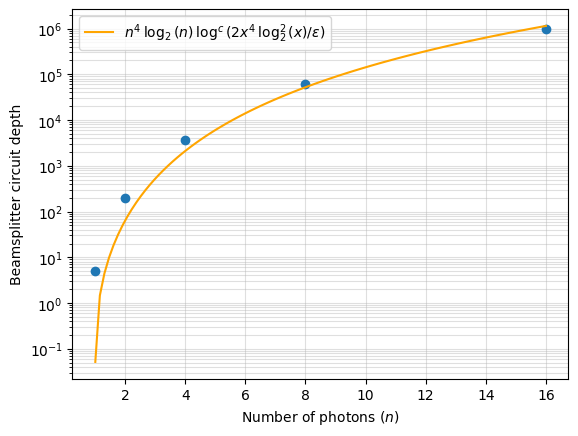}
    \caption{Circuit depths for a randomly selected two-mode interferometer for various numbers of photons. The orange curve is placed as a visual aid to show the approximate trend of the data points; Its vertical coordinates are equal to $\mathcal{O}\Big(n^4 \log_2(n) \: \log^c(2 n^4 \log_2^2(n)/\epsilon)\Big)$ where $n$ is the photon number, $\epsilon$ is an error term (assumed to be $1\times 10^{-15}$ since the numerical errors reported in section \ref{sec:results} have the same order of magnitude) and $c \approx 0.377$ (which was selected using non-linear least squares regression)}.
    \label{fig:circuit_scaling}
\end{figure*}

In order to calculate the depth of \texttt{Aquinas}'s circuits with respect to $m$ and $n$, it is sufficient to calculate the depths of the beamsplitter circuits with respect to $n$. 
This is because every circuit with $m > 2$ modes is composed of multiple beamsplitter circuits arranged in a predictable pattern (Recall figure \ref{fig:beamsplitter_decomp}).
A scatterplot of these depths are presented in fig.\ \ref{fig:circuit_scaling}.

\begin{widetext}
\begin{algorithm}[tbp]
 \DontPrintSemicolon
 \SetKwInOut{Input}{Inputs}
 \SetKwInOut{Output}{Output}
 \SetKwFunction{shortestPath}{best\_path\_wrt\_coherence}
 \Input{
 \begin{itemize}
     \item $n$ (The maximum photon number)
     \item $U$ (An $m \times m$ unitary matrix describing how creation operators are transformed by an $m$ mode interferometer)
     \item \texttt{sparsity\_optimization} == True (If true, we perform an additional `padding' step that slightly improves the sparsity of the circuit unitary when $n$ is not a power of two.)
 \end{itemize}
 }
 \Output{A quantum circuit that simulates the interferometer for any photon number state containing up to (and including) $n$ photons.
 }
 \BlankLine
 \Begin{
     Decompose $U$ to obtain $m(m-1)/2$ many $2\times 2$ unitary matrices. Remember their positions. \;
 \ForEach{$2 \times 2$ unitary $B_i$}{ Calculate the corresponding Hamiltonian $H_i$ (See section \ref{sec:worked_example} for an example) \;
 \# Calculate $H^{(n)}_i$: \;
 \If{sparsity\_optimization == False}{
    n := \texttt{next\_highest\_power\_of\_two}(n) \;
    Replace unbounded creation and annihilation operators $(\hat a^\dagger, \:\hat a)$ with truncated counterparts $(\hat a^{ \dagger (n)}, \:\hat a^{(n)})$ \;
    }
 \Else{
        Replace $(\hat a^\dagger, \:\hat a)$ with $(\hat a^{ \dagger (n)}, \:\hat a^{(n)})$ \;
        Pad $\hat a^{(n)^\dagger}$ and $\hat a^{(n)}$ with zeros until the shapes of each are a power of two. (See eq.\ \ref{eq:padding}) \;
        }
Take the exponential of $H^{(n)}_i$ to obtain $U^{(n)}_i$ \;
Convert $U^{(n)}_i$ into a quantum circuit (We use the \texttt{Qiskit.transpile()} method \cite{qiskit2024} at \texttt{optimization\_level} = 2 to synthesise circuits from the gateset $\{CNOT, U_3\}$)
 }
Assemble the circuits as per the original decomposition
 }
\caption{Pseudocode for the algorithm used by the \texttt{Aquinas} software.
} \label{alg:aquinas}
\end{algorithm}
\end{widetext}



\newpage
\newpage

\section{Conclusion and outlook}

In this work, we looked at approaches for simulating linear optics in first and second quantization pictures.
We presented a divide-and-conquer strategy in the second quantization picture based on decomposing the interferometer into beamsplitters and phase shifters, which are more easily synthesised into circuits.
We wrote software to construct interferometer circuits according to this method, and found that the asymptotic circuit size and depth were $\mathcal{O}(m \log(n))$ and $\mathcal{O}\Big(n^4 \log_2(n) \: \textrm{polylog}(n^4 / \epsilon) \Big)$ respectively.
Finally, we validated our efforts by presenting two numerical experiments in fig.\ \ref{fig:random_3x3}.

Although we initially hoped that this \textit{divide-and-conquer} method might be able to generate feasible circuits for non-trivial instances of boson sampling (upwards of $n = 30$ and $m = \mathcal{O}(n^2)$ \cite{Aaronson_11}), it is clear from our numerical data (fig.\ \ref{fig:circuit_scaling}) that the circuit depths for individual beamsplitters are prohibitively large; A 16 photon beamsplitter for example requires a depth of around $10^6$ (see fig.\ \ref{fig:circuit_scaling}).
Consequently, it is unlikely that \textit{divide-and-conquer} will see any practical use, even with further optimizations.


One promising direction for further research is to simulate in the first quantization picture instead. As mentioned in section \ref{sec:background}, the main difficulty with this approach is the preparation of symmetric states.
Recently, Berry et.\ al.\ proposed various methods for efficiently preparing \textit{anti-symmetric states} \cite{Berry_18}. 
Using a sorting network called an \textit{odd-even mergesort}, they indicate that it is possible to prepare an $m$ mode, $n$ particle anti-symmetric state using $\mathcal{O}(m \log^c m \log n)$ ancilla qubits and a circuit depth of $\mathcal{O}(\log^c m \log\log n)$.
With further work, it is possible their techniques could be used to generate symmetric states.


In this work, we have limited out discussion to the circuit model of quantum computation. Whether or not there is an advantage in considering a measurement-based framework is an open question. Our work can be used as a first step in this direction by noting that \texttt{Aquinas} circuits can be translated into equivalent graphs using the \texttt{Jabilizer} package by Vijayan et.\ al.\ \cite{Jabilizer}.

\section{Acknowledgements}

We thank Prof.\ Gavin Brennen and Dr.\ Peter Rhode for helpful discussions.
The lead author extends a special thanks to our resident computer wizard Dr.\ Alan Robertson for helping fix a particularly nasty bug.

\clearpage
\bibliography{paperbib.bib}

\begin{thebibliography}{25}
\providecommand{\natexlab}[1]{#1}
\providecommand{\url}[1]{\texttt{#1}}
\expandafter\ifx\csname urlstyle\endcsname\relax
  \providecommand{\doi}[1]{doi: #1}\else
  \providecommand{\doi}{doi: \begingroup \urlstyle{rm}\Url}\fi

\bibitem[Kok et~al.(2007)Kok, Munro, Nemoto, Ralph, Dowling, and Milburn]{Kok_07}
Pieter Kok, W.~J. Munro, Kae Nemoto, T.~C. Ralph, Jonathan~P. Dowling, and G.~J. Milburn.
\newblock Linear optical quantum computing with photonic qubits.
\newblock \emph{Rev. Mod. Phys.}, 79:\penalty0 135--174, Jan 2007.
\newblock \doi{10.1103/RevModPhys.79.135}.
\newblock URL \url{https://link.aps.org/doi/10.1103/RevModPhys.79.135}.

\bibitem[Aaronson and Arkhipov(2011)]{Aaronson_11}
Scott Aaronson and Alex Arkhipov.
\newblock The computational complexity of linear optics.
\newblock In \emph{Proceedings of the Forty-Third Annual ACM Symposium on Theory of Computing}, STOC '11, page 333–342, New York, NY, USA, 2011. Association for Computing Machinery.
\newblock ISBN 9781450306911.
\newblock \doi{10.1145/1993636.1993682}.
\newblock URL \url{https://doi.org/10.1145/1993636.1993682}.

\bibitem[Brod et~al.(2019)Brod, Galvão, Crespi, Osellame, Spagnolo, and Sciarrino]{Brod_19}
Daniel~J. Brod, Ernesto~F. Galvão, Andrea Crespi, Roberto Osellame, Nicolò Spagnolo, and Fabio Sciarrino.
\newblock Photonic implementation of boson sampling: a review.
\newblock \emph{Advanced Photonics}, 1\penalty0 (11):\penalty0 034001, 3 2019.
\newblock URL \url{https://www.researching.cn/articles/OJ1d41e4f96f414592}.

\bibitem[Madsen et~al.(2022)Madsen, Laudenbach, Askarani, Rortais, Vincent, Bulmer, Miatto, Neuhaus, Helt, Collins, Lita, Gerrits, Nam, Vaidya, Menotti, Dhand, Vernon, Quesada, and Lavoie]{Madsen2022}
Lars~S. Madsen, Fabian Laudenbach, Mohsen~Falamarzi. Askarani, Fabien Rortais, Trevor Vincent, Jacob F.~F. Bulmer, Filippo~M. Miatto, Leonhard Neuhaus, Lukas~G. Helt, Matthew~J. Collins, Adriana~E. Lita, Thomas Gerrits, Sae~Woo Nam, Varun~D. Vaidya, Matteo Menotti, Ish Dhand, Zachary Vernon, Nicolás Quesada, and Jonathan Lavoie.
\newblock Quantum computational advantage with a programmable photonic processor.
\newblock \emph{Nature}, 606\penalty0 (7912):\penalty0 75–81, June 2022.
\newblock ISSN 1476-4687.
\newblock \doi{10.1038/s41586-022-04725-x}.
\newblock URL \url{http://dx.doi.org/10.1038/s41586-022-04725-x}.

\bibitem[Arrazola and Bromley(2018)]{Arrazola_18}
Juan~Miguel Arrazola and Thomas~R. Bromley.
\newblock Using gaussian boson sampling to find dense subgraphs.
\newblock \emph{Phys. Rev. Lett.}, 121:\penalty0 030503, Jul 2018.
\newblock \doi{10.1103/PhysRevLett.121.030503}.
\newblock URL \url{https://link.aps.org/doi/10.1103/PhysRevLett.121.030503}.

\bibitem[Brádler et~al.(2021)Brádler, Friedland, Izaac, Killoran, and Su]{Bradler_21}
Kamil Brádler, Shmuel Friedland, Josh Izaac, Nathan Killoran, and Daiqin Su.
\newblock Graph isomorphism and gaussian boson sampling.
\newblock \emph{Special Matrices}, 9\penalty0 (1):\penalty0 166--196, 2021.
\newblock \doi{doi:10.1515/spma-2020-0132}.
\newblock URL \url{https://doi.org/10.1515/spma-2020-0132}.

\bibitem[Singh et~al.(2024)Singh, Muraleedharan, Fu, Cheng, Newton, Rohde, and Brennen]{Singh_2024}
Deepesh Singh, Gopikrishnan Muraleedharan, Boxiang Fu, Chen-Mou Cheng, Nicolas~Roussy Newton, Peter~P. Rohde, and Gavin~K. Brennen.
\newblock Proof-of-work consensus by quantum sampling, 2024.
\newblock URL \url{https://arxiv.org/abs/2305.19865}.

\bibitem[Heurtel et~al.(2023)Heurtel, Mansfield, Senellart, and Valiron]{Heurtel2023}
Nicolas Heurtel, Shane Mansfield, Jean Senellart, and Benoît Valiron.
\newblock Strong simulation of linear optical processes.
\newblock \emph{Computer Physics Communications}, 291:\penalty0 108848, October 2023.
\newblock ISSN 0010-4655.
\newblock \doi{10.1016/j.cpc.2023.108848}.
\newblock URL \url{http://dx.doi.org/10.1016/j.cpc.2023.108848}.

\bibitem[Aaronson and Brod(2016)]{Aaronson_16}
Scott Aaronson and Daniel~J. Brod.
\newblock Bosonsampling with lost photons.
\newblock \emph{Phys. Rev. A}, 93:\penalty0 012335, Jan 2016.
\newblock \doi{10.1103/PhysRevA.93.012335}.
\newblock URL \url{https://link.aps.org/doi/10.1103/PhysRevA.93.012335}.

\bibitem[Oszmaniec and Brod(2018)]{Oszmaniec2018}
Michał Oszmaniec and Daniel~J Brod.
\newblock Classical simulation of photonic linear optics with lost particles.
\newblock \emph{New Journal of Physics}, 20\penalty0 (9):\penalty0 092002, September 2018.
\newblock ISSN 1367-2630.
\newblock \doi{10.1088/1367-2630/aadfa8}.
\newblock URL \url{http://dx.doi.org/10.1088/1367-2630/aadfa8}.

\bibitem[Moylett et~al.(2019)Moylett, García-Patrón, Renema, and Turner]{Moylett2019}
Alexandra~E Moylett, Raúl García-Patrón, Jelmer~J Renema, and Peter~S Turner.
\newblock Classically simulating near-term partially-distinguishable and lossy boson sampling.
\newblock \emph{Quantum Science and Technology}, 5\penalty0 (1):\penalty0 015001, November 2019.
\newblock ISSN 2058-9565.
\newblock \doi{10.1088/2058-9565/ab5555}.
\newblock URL \url{http://dx.doi.org/10.1088/2058-9565/ab5555}.

\bibitem[Leone()]{aquinas}
Hudson Leone.
\newblock Aquinas.
\newblock \url{https://github.com/FalafelGood/Aquinas}.
\newblock Version: 1.0, Accessed: 2024-08-30.

\bibitem[S and Y.(2008)]{Scheel_08}
Scheel S and Buhmann~S. Y.
\newblock Macroscopic quantum electrodynamics-concepts and applications.
\newblock \emph{Acta Physica Slovaca}, 58:\penalty0 675--809, October 2008.
\newblock ISSN 0323-0465.

\bibitem[Moylett and Turner(2018)]{Moylett_18}
Alexandra~E. Moylett and Peter~S. Turner.
\newblock Quantum simulation of partially distinguishable boson sampling.
\newblock \emph{Phys. Rev. A}, 97:\penalty0 062329, Jun 2018.
\newblock \doi{10.1103/PhysRevA.97.062329}.
\newblock URL \url{https://link.aps.org/doi/10.1103/PhysRevA.97.062329}.

\bibitem[Bacon et~al.(2006)Bacon, Chuang, and Harrow]{Bacon_06}
Dave Bacon, Isaac~L. Chuang, and Aram~W. Harrow.
\newblock Efficient quantum circuits for schur and clebsch-gordan transforms.
\newblock \emph{Phys. Rev. Lett.}, 97:\penalty0 170502, Oct 2006.
\newblock \doi{10.1103/PhysRevLett.97.170502}.
\newblock URL \url{https://link.aps.org/doi/10.1103/PhysRevLett.97.170502}.

\bibitem[Sawaya et~al.(2020)Sawaya, Menke, Kyaw, Johri, Aspuru-Guzik, and Guerreschi]{Sawaya2020}
Nicolas P.~D. Sawaya, Tim Menke, Thi~Ha Kyaw, Sonika Johri, Alán Aspuru-Guzik, and Gian~Giacomo Guerreschi.
\newblock Resource-efficient digital quantum simulation of d-level systems for photonic, vibrational, and spin-s hamiltonians.
\newblock \emph{npj Quantum Information}, 6\penalty0 (1), June 2020.
\newblock ISSN 2056-6387.
\newblock \doi{10.1038/s41534-020-0278-0}.
\newblock URL \url{http://dx.doi.org/10.1038/s41534-020-0278-0}.

\bibitem[Clements et~al.(2016)Clements, Humphreys, Metcalf, Kolthammer, and Walmsley]{Clements_16}
William~R. Clements, Peter~C. Humphreys, Benjamin~J. Metcalf, W.~Steven Kolthammer, and Ian~A. Walmsley.
\newblock Optimal design for universal multiport interferometers.
\newblock \emph{Optica}, 3\penalty0 (12):\penalty0 1460--1465, Dec 2016.
\newblock \doi{10.1364/OPTICA.3.001460}.
\newblock URL \url{https://opg.optica.org/optica/abstract.cfm?URI=optica-3-12-1460}.

\bibitem[Note1()]{Note1}
Note1.
\newblock The reason this unitary was chosen is because it can be expressed as a time-independent Hamiltonian. It is worth noting though that the Clements unitary given in eq.\ \ref {eq:clements_unitary} can be written as a \protect \textit {product} of time-independent Hamiltonians (See eq.\ 4.13 in ref \cite {Arrazola2020}). This is needlessly complex for the sake of our example, which is why we use the eq.\ \ref {eq:simple_unitary} unitary instead.

\bibitem[Nielsen and Chuang(2012)]{Nielsen2012}
Michael~A. Nielsen and Isaac~L. Chuang.
\newblock \emph{Quantum Computation and Quantum Information: 10th Anniversary Edition}.
\newblock Cambridge University Press, June 2012.
\newblock ISBN 9780511976667.
\newblock \doi{10.1017/cbo9780511976667}.
\newblock URL \url{http://dx.doi.org/10.1017/CBO9780511976667}.

\bibitem[Hurwitz(1897)]{Hurwitz1897}
A.~Hurwitz.
\newblock über die erzeugung der invarianten durch integration.
\newblock \emph{Nachrichten von der Gesellschaft der Wissenschaften zu Göttingen, Mathematisch-Physikalische Klasse}, 1897:\penalty0 71--2, 1897.
\newblock URL \url{http://eudml.org/doc/58378}.

\bibitem[Reck et~al.(1994)Reck, Zeilinger, Bernstein, and Bertani]{Reck_94}
Michael Reck, Anton Zeilinger, Herbert~J. Bernstein, and Philip Bertani.
\newblock Experimental realization of any discrete unitary operator.
\newblock \emph{Phys. Rev. Lett.}, 73:\penalty0 58--61, Jul 1994.
\newblock \doi{10.1103/PhysRevLett.73.58}.
\newblock URL \url{https://link.aps.org/doi/10.1103/PhysRevLett.73.58}.

\bibitem[Javadi-Abhari et~al.(2024)Javadi-Abhari, Treinish, Krsulich, Wood, Lishman, Gacon, Martiel, Nation, Bishop, Cross, Johnson, and Gambetta]{qiskit2024}
Ali Javadi-Abhari, Matthew Treinish, Kevin Krsulich, Christopher~J. Wood, Jake Lishman, Julien Gacon, Simon Martiel, Paul~D. Nation, Lev~S. Bishop, Andrew~W. Cross, Blake~R. Johnson, and Jay~M. Gambetta.
\newblock Quantum computing with {Q}iskit, 2024.

\bibitem[Berry et~al.(2018)Berry, Kieferová, Scherer, Sanders, Low, Wiebe, Gidney, and Babbush]{Berry_18}
Dominic~W. Berry, Mária Kieferová, Artur Scherer, Yuval~R. Sanders, Guang~Hao Low, Nathan Wiebe, Craig Gidney, and Ryan Babbush.
\newblock Improved techniques for preparing eigenstates of fermionic hamiltonians.
\newblock \emph{npj Quantum Information}, 4\penalty0 (1), May 2018.
\newblock ISSN 2056-6387.
\newblock \doi{10.1038/s41534-018-0071-5}.
\newblock URL \url{http://dx.doi.org/10.1038/s41534-018-0071-5}.

\bibitem[Krishnan~Vijayan et~al.(2024)Krishnan~Vijayan, Paler, Gavriel, Myers, Rohde, and Devitt]{Jabilizer}
Madhav Krishnan~Vijayan, Alexandru Paler, Jason Gavriel, Casey~R Myers, Peter~P Rohde, and Simon~J Devitt.
\newblock Compilation of algorithm-specific graph states for quantum circuits.
\newblock \emph{Quantum Science and Technology}, 9\penalty0 (2):\penalty0 025005, February 2024.
\newblock ISSN 2058-9565.
\newblock \doi{10.1088/2058-9565/ad1f39}.
\newblock URL \url{http://dx.doi.org/10.1088/2058-9565/ad1f39}.

\bibitem[Arrazola(2020)]{Arrazola2020}
I.~Arrazola.
\newblock \emph{{Design of Light-Matter Interactions for Quantum Technologies}}.
\newblock PhD thesis, U. Basque Country, Leioa, 2020.

\end{thebibliography}

\end{document}